\documentclass[12pt]{article}
\usepackage{epsfig,rotating,setspace,latexsym,amsmath,epsf,amssymb,amsfonts,bm,bbm,theorem,cite,caption,subcaption,enumerate,longtable,accents,theoremref,mathtools,mathrsfs}
\usepackage{graphicx,epsf,authblk,epstopdf,url,color,multirow}
\usepackage{soul}
\usepackage[ruled,vlined]{algorithm2e}

\newtheorem{theorem}{Theorem}

\newtheorem{lemma}{Lemma}
\newenvironment{Proof}[1]{\medskip\par\noindent{\bf Proof:\,}\,#1}{{\mbox{\,$\blacksquare$}\par}}
\newcounter{example}

\allowdisplaybreaks

\begin{document}

\title{Dynamic Infection Spread Model Based Group Testing}
\author{Batuhan Arasli \qquad Sennur Ulukus\\
	\normalsize Department of Electrical and Computer Engineering\\
	\normalsize University of Maryland, College Park, MD 20742 \\
	\normalsize {\it barasli@umd.edu} \qquad {\it ulukus@umd.edu}}

\date{}
\maketitle

\vspace*{-1cm}

\begin{abstract}
We study a dynamic infection spread model, inspired by the discrete time SIR model, where infections are spread via non-isolated infected individuals. While infection keeps spreading over time, a limited capacity testing is performed at each time instance as well. In contrast to the classical, static, group testing problem, the objective in our setup is not to find the minimum number of required tests to identify the infection status of every individual in the population, but to \emph{control the infection spread by detecting and isolating the infections over time by using the given, limited number of tests}. In order to analyze the performance of the proposed algorithms, we focus on the mean-sense analysis of the number of individuals that remain non-infected throughout the process of controlling the infection. We propose two dynamic algorithms that both use given limited number of tests to identify and isolate the infections over time, while the infection spreads. While the first algorithm is a dynamic randomized individual testing algorithm, in the second algorithm we employ the group testing approach similar to the original work of Dorfman. By considering weak versions of our algorithms, we obtain lower bounds for the performance of our algorithms. Finally, we implement our algorithms and run simulations to gather numerical results and compare our algorithms and theoretical approximation results under different sets of system parameters.
\end{abstract}

\section{Introduction}
The group testing idea, introduced by Dorfman in his seminal work \cite{dorfman1943}, is an efficient approach to the detection of the prevalence of a certain infection in the test samples of a group of individuals. The group testing approach is based on the idea of dividing the individuals into groups, mixing the collected test samples within each group and testing those mixed samples. This way, a negative test result implies that every test sample included in that mixed sample is negative, while a positive test result implies that there is at least one positive sample in the mixed sample: each test acts like a logical OR operation of the test samples included in the mixed sample.

Dorfman's original algorithm divides the population into disjoint groups and performs group tests by mixing the samples within each group. Subsequently, depending on the test results, positive groups are further tested individually to identify the status of every individual in the population. After Dorfman's work in \cite{dorfman1943}, various adaptive (tests performed in multiple steps) and non-adaptive (tests performed in a single step) group testing algorithms have been proposed, capacity of the group testing problem has been studied for a variety of system models and family of algorithms, and extended analyses have been conducted for different regimes for the total number of infections in the population \cite{binarysplittingorig,hwang_binary,hwang_disjunct,RUSZINKO,nonadaptive_bounds,adaptivecapacity,mazumdar_nonadaptive,scarlett_noisynonadaptive,wu_partition,wang_combquant,combinatorial_gt,bornagain_mac,ddintroduction,sharper,scarlettbook,cai_noisy,karimi_irregularsparsegraph,inan_optimalityks,johnson_nearconstant,allemann,scarlett_noisyadaptivebounds,atia_saligrama_first}.

Most of the classical group testing problems have considered either combinatorial or i.i.d.~probabilistic settings. Out of $n$ individuals, in the combinatorial setting, some fixed number of individuals, $k$, are assumed to be infected and the problem is to identify the infected set that is uniformly randomly selected from the set of all size $k$ subsets of $n$ individuals with the minimum number of tests, while in the i.i.d.~probabilistic group testing models, each individual is assumed to be randomly infected, with some fixed probability, independently. In the classical system models, under various assumptions, it is proven that the advantage of group testing over individual testing is considerable mostly in the scenarios where the infection prevalence rate in the population is not high \cite{sharper,RUSZINKO,scarlettbook} and the advantage gets diminished as the infection prevalence rate gets higher within the population. More recently, there has been an increasing focus on modified system models, where practical considerations on the system models have significantly improved the performance of the group testing systems. \cite{prior_stats} considers a probabilistic model where each individual is randomly infected with an unequal probability, independently. \cite{doger2021GroupTW} studies non-identical infection probability distributions and proposes a novel adaptive group testing algorithm. \cite{correlated_bio,lincorrelated} consider correlation between infection status of individuals, and \cite{diggavicommunity,diggavioverlap,ayferozgurcommunity,arasli2021group} further model the correlations by considering community structures among the individuals. 

Although these extended system models are practical and resemble real-life scenarios, in reality, the testing and infection identification processes are dynamic in nature, especially for contagious diseases. Instead of a static, single-shot identification procedure as in the classical group testing setup, testing and identification need to be done over a long time period while the infection keeps spreading and the infection status of the individuals are dynamically changing. Furthermore, rather than minimizing the required number of tests for perfect (or near perfect with vanishing error) identification of the infection status of a population as in the classical group testing setups, in most practical scenarios, the limit on the testing capacity might be fixed and the objective might be to identify as many infections as possible with the limited testing capacity. 

Within this context, \cite{aldridgeisolate} considers a limited identification scenario in a classical setting, addressing the limited test capacity with partial identification cases for real-life scenarios. In a continuous setting and without focusing on explicit testing algorithms, but approaching the problem as a control system with controlled variable testing rate, \cite{acemoglu2021optimal} considers a dynamic testing infection spread model based on the well-known SIR model \cite{sirmodel2020}. However, the first works that focus on the testing of individuals in a dynamic setup with an emphasis on group testing are \cite{dynamicgtdiggavi,dynamicgtdiggavi2}, where the authors consider a delay in test results and minimize the number of required tests to identify the infection status of the population at each time instance, stating an equivalence to the classical static group testing problem. Despite the fact that the tests are performed to identify the infection status of everyone in the population, new infections are introduced into the system, due to the assumed delay in test results, i.e., during the time between performing the tests and getting the test results. In a more recent work, \cite{doger2022dynamical} considers a scenario where the tests are done in a dynamic manner, with a focus on two-step Dorfman testing with delay and disease spread between the two steps, in a discrete SIR model with a community structure; the authors in \cite{doger2022dynamical} consider quarantining the possible infections, and analyze the trade-off between quarantine and test costs. 

In this paper, we consider dynamic testing algorithms over discrete time for a dynamic infection spread model with fixed, limited testing capacity at each time instance, where a full identification is not possible. In our system, test results are available immediately, and thus, the disease spread is not due to the delay between applying tests and receiving test results, but rather due to the limited testing capacity at each time instance. We follow a dynamic infection spread model, similar to the infection spread model in \cite{dynamicgtdiggavi}, which is inspired by the well-known SIR model where the individuals are divided into three groups: susceptible individuals (S), non-isolated infections (I) and isolated infections (R), i.e., recovered individuals in classical SIR model. We do not assume a community structure in our system. We initialize our system by introducing the initial infections, and after that, at each time instance, infection is spread by infected non-isolated individuals to the susceptible individuals. Meanwhile, at each time instance, after the infection spread phase, the testing phase is performed, where a limited number of $T$ tests are performed to detect a number of infections in the system. In our system, the objective is not to minimize the number of required tests to identify everyone at each time instance, but to control the infection spread either as soon as possible or with minimum number of people that got infected throughout the process, by using the given, limited, testing capacity $T$ at each time instance.

In this paper, we analyze the mean-sense performance of our system, i.e., the expected values of the number of susceptible individuals, non-isolated and isolated infections over time, which are random processes. For \emph{symmetric and converging algorithms}, we state a general analytical result for the expected number of susceptible individuals in the system when the infection is brought under control, which is the time when there is no non-isolated infection left in the system. We present two dynamic algorithms: dynamic individual testing and dynamic Dorfman type group testing algorithm. We provide weak versions of these two algorithms and use our general result to obtain a lower bound on the expected number of susceptible individuals when the infection is under control. Finally, we run simulations to get numerical results of our proposed algorithms for different sets of parameters.

\section{System Model} \label{sec2}
We consider a population of $n$ individuals whose infection status change over time. The time dimension $t$ is discrete in our system, i.e., $t \in \{0, 1, 2, \ldots\}$. Similar to the classical discrete SIR model, the population consists of three distinct subgroups: susceptible individuals who are not infected but can get infected by infected individuals (S), infected individuals who can infect the susceptible individuals (I), and isolated individuals who were infected, have been detected via performed tests and isolated indefinitely (R)\footnote{These are called recovered (R) individuals in the SIR model; we call them isolated individuals. As they are isolated indefinitely, they are recovered eventually.}. Let $U_{i}(t)$ denote the infection status of individual $i$ at time $t$, where 1 represents being infected, 0 represents not being infected and 2 represents being isolated. At the beginning ($t=0$), we introduce the initial infections in the system, independently with probability $p$, where $U_{i}(0)$ is a Bernoulli random variable with parameter $p$. Random variables $U_i(0)$ are mutually independent for $i \in [n]$. Let $\alpha (t)$ denote the number of susceptible individuals at time $t$, $\lambda (t)$ denote the number of non-isolated infected individuals at time $t$ and $\gamma (t)$ denote the number of isolated individuals at time $t$. Starting from $t=1$, each time instance consists of two phases: infection spread phase and testing phase, in the respective order.

\paragraph{Infection Spread Phase:}Infected individuals spread the infection to the susceptible members of the population. At each time instance, starting from $t=1$, the infection spreads independently across the individuals: Each infected individual can infect each susceptible individual with probability $q$, independent across both infected individuals and susceptible individuals. Isolated individuals cannot infect others and their infection status cannot change after they are isolated. Thus, probability of the event that individual $i$ gets infected by another individual $j$ at time $t \geq 1$ is equal to $qP\left(U_j(t-1)=1, U_i(t-1)=0\right)$ for $i,j \in [n]$.

\paragraph{Testing Phase:}At each time instance starting from $t=1$, $T$ tests can be performed to the individuals. Note that the testing capacity $T$ is a given parameter and thus, in contrast to the classical group testing systems, we do not seek to minimize the number of performed tests for full identification of the infection status of the population but aim to efficiently perform $T$ tests at each time instance to identify and isolate as many infections as possible to control the infection spread.  Here, performed tests can be group tests, and we define the $T \times n$ binary test matrices, $\bm{X}(t)$, which specify the pooling scheme for the tests at each time $t$. For each time instance $t \geq 1$, we have the test result vectors $y(t)$, which are equal to
\begin{align}
	y_i(t) = \bigvee _ {j \in [n]} \bm{X}_{ij}(t) \mathbbm{1}_{\{U_{j}(t)=1\}}, \quad i \in [T]
\end{align}
where $y_i(t)$ denotes the $i$th test result at time $t$, $\bm{X}_{ij}(t)$ denotes the $i$th row, $j$th column of the test matrix $\bm{X}(t)$.

Note that, since the previous test matrices and test results are available while designing these test matrices, $\bm{X}(t)$ can depend on the previous test results $y(t')$ for $t'<t$. We assume that when tests are performed at some time instance $t'$, the test results $y(t')$ will be available before the infection spread phase at time $t'+1$. Thus, after the test results are available, detected infections are isolated immediately, i.e., if the $i$th individual is detected to be infected during the testing phase at time $t'$, then $U_i(t')=2$. Recall that, after an infected individual is isolated at some time $t'$, they cannot infect others at times greater than $t'$ and their infection status cannot change, i.e., $U_{i}(t) = 2$ for $t \geq t'$.

\paragraph{Testing Policy:}A testing policy $\pi$ is an algorithm that specifies how to allocate the given testing capacity $T$ for each time instance, until the infection is under control. We define $\bar{t}$ to be the time when $U_i(\bar{t}) \neq 1$ for all individuals $i \in [n]$ for the first time and we say that the infection is under control at $\bar{t}$. Note that, after $\bar{t}$, infection status of the individuals cannot change and the steady state is achieved: They are either isolated ($U_i(t)=2$) or non-infected ($U_i(t)=0$). Since we do not consider re-entries of recoveries to the population, the infection spread is under control when all infections in the system are isolated. Otherwise, infection may keep spreading to the susceptible individuals by the non-detected infections.

\paragraph{Performance Metrics:} The main objective is to bring the infection spread under control by detecting and isolating each infected individual by performing at most $T$ tests at each time instance. Note that, meanwhile, infection keeps spreading and thus, detecting the infection status of an individual to be negative does not imply that they are identified for the rest of the process; they can get infected in the later time instances. As defined, $\bar{t}$ is the time that the infection is under control and when the system is reached that state, further testing of the individuals is unnecessary. Therefore, there are two metrics to measure the performance of a testing policy $\pi$: The time $\bar{t}$ when the infection is brought under control and the total number of isolated individuals when the infection is under control. While comparing the performances of the testing policies, earlier infection control time $\bar{t}$ and less number of total infections at the time of infection control $\gamma(\bar{t})$ are favored. Proposed algorithms may not simultaneously improve both metrics: One policy may bring the infection spread under control fast (i.e., low $\bar{t}$) but may result in high number of total infections (i.e., high $\gamma(\bar{t})$) while another policy may bring the infection spread under control slowly but with lower number of total infections.

\section{Proposed Algorithms and Analysis} \label{sec3}
In this section, we propose two algorithms and analyze their performances. First algorithm does not utilize the group testing approach and it is based on the idea of dynamically and individually testing the population. Second algorithm consists of a group testing approach at each time instance, similar to the original idea of Dorfman \cite{dorfman1943} in a dynamic setting. Before stating these two algorithms and further analyzing their performances individually, we first state general results.

\paragraph{Symmetric and Converging Dynamic Testing Algorithms:} In our analysis, we focus on \emph{symmetric and converging dynamic testing algorithms}, which satisfy the \emph{symmetry} criterion,
\begin{align}
    P(U_i(t)=k) = P(U_j(t)=k) , \quad i,j \in [n] \quad k \in \{0,1,2\} \quad t \geq 0 \label{symalgcr}
\end{align}
and \emph{convergence} criterion,
\begin{align}
    \lim_{t \rightarrow \infty}P(U_i(t)=1) = o(1/n), \quad i \in [n] \label{convalgcr}
\end{align}
Furthermore, we assume that the probability of an individual not being identified in the tests at time $t$, denoted by $p'(t)$, only depends on the testing capacity $T$, $\alpha (t)$, $\lambda (t)$ and $\gamma (t)$. Note that, $\alpha (t)+\lambda (t)+\gamma (t) = n$ for all time instances $t$.

\paragraph{Infection Spread Probability:} We consider $q=o(1/n)$ for the infection spread probability $q$. This is a practical assumption since $q$ is the probability of the event of infection spread that is realized independently for every element of the set product of the infected individuals and susceptible individuals, at each time instance.

We analyze the long term behavior of the system in the mean sense, i.e., we focus on the terms $E[\alpha(t)]$, $E[\lambda (t)]$ and $E[\gamma (t)]$ when $t$ is large enough.

\begin{lemma} \label{lemma1}
When a symmetric and converging dynamic testing algorithm is implemented, $\lim_{t \rightarrow \infty}E[\lambda (t)] = o(1)$ and thus, the system approaches to steady state, in the mean sense.
\end{lemma}

\begin{Proof}
Note that all three system functions $\alpha (t)$, $\lambda (t)$ and $\gamma (t)$ can be written as the summation of $n$ indicator functions
\begin{align}
    E[\lambda (t)] &= E\left[\sum \limits_{i = 1}^{n}\mathbbm{1}_{\{U_i(t)=1\}}\right] \\
    &= \sum \limits_{i = 1}^{n} E\left[\mathbbm{1}_{\{U_i(t)=1\}}\right] \\
    &= \sum \limits_{i = 1}^{n} P(U_i(t)=1)\\
    &= n P(U_i(t)=1)
\end{align}
which results in $\lim_{t \rightarrow \infty}E[\lambda (t)] = no(1/n)$, due to converging algorithm assumption \eqref{convalgcr}, which is equal to $o(1)$.
\end{Proof}

Note that, when the system reaches a state where $\lambda (t) = 0$, then there will not be further change of the infection status of the individuals, i.e., the infection will be under control. The following lemma is useful for the justification of the mean sense analysis of our system.

\begin{lemma}
When a symmetric and converging dynamic testing algorithm is implemented, we have $\lim_{t \rightarrow \infty}P(\lambda (t) > \epsilon) = o(1)$ for arbitrarily small, constant, $\epsilon \in \mathbbm{R}$.
\end{lemma}

\begin{Proof}
Since $\lambda(t) \geq 0$ for all $t \geq 0$, we can apply Markov's inequality,
\begin{align}
    \lim_{t \rightarrow \infty} P(\lambda (t) > \epsilon) &\leq \lim_{t \rightarrow \infty} \frac{E[\lambda (t)]}{\epsilon} \label{lem2l1}\\
    &= o(1) \label{lem2l2}
\end{align}
where \eqref{lem2l1} follows from the fact that $P(\lambda (t) > \epsilon) \leq  \frac{E[\lambda (t)]}{\epsilon}$ for all $t \geq 0$, and \eqref{lem2l2} follows from the result of Lemma~\ref{lemma1}.
\end{Proof}

The focus of our analysis is to give a lower bound for the number of susceptible individuals (who have never got infected throughout the process) when the infection is brought under control, in the mean sense. In order to analyze the long term behavior of $E[\alpha(t)]$, we have to analyze the long term behavior of $P(U_i(t)=0)$. A direct calculation of this probability is not analytically tractable, however, by conditioning on $\lambda(t-1)$, we give a recursive asymptotic calculation. Before stating the recursive relation, we first prove a lemma that will be useful.

\begin{lemma} \label{lemma33}
For $q=o(1/n)$ and for all $t \geq 0$, we have
\begin{align}
     cov\left(P(U_i(t)=0|\lambda (t)),(1-q)^{\lambda (t)}\right) \approx 0
\end{align}
\end{lemma}

\begin{Proof}
For the proof, we use the covariance inequality, i.e., $|cov(X,Y)| \leq \sqrt{var(X)var(Y)}$ which is a direct application of the Cauchy-Schwarz inequality, applied to the random variables $X-E[X]$ and $Y-E[Y]$. Using the covariance inequality, we have
\begin{align}
     |cov\left(P(U_i(t)=0|\lambda (t)),(1-q)^{\lambda (t)}\right)| &\leq \sqrt{var(P(U_i(t)=0|\lambda (t)))var((1-q)^{\lambda (t)})} \\
     &\leq \sqrt{var((1-q)^{\lambda (t)})} \label{lem33l1}\\ 
     &= \sqrt{E[(1-q)^{2\lambda (t)}]-(E[(1-q)^{\lambda (t)}])^2}\\
     &\approx \sqrt{(1-q)^{E[2\lambda (t)]}-((1-q)^{E[\lambda (t)]})^2}\label{lem33l2}\\
     &= 0
\end{align}
where \eqref{lem33l1} follows from the fact that the random variable $P(U_i(t)=0|\lambda (t))$ is bounded above by 1 and below by 0, and \eqref{lem33l2} follows from the linear approximation of the function $(1-q)^x$ for small $q=o(1/n)$ and $\lambda (t)$ that is bounded above by $n$.
\end{Proof}

\begin{lemma} \label{lemma3}
When a symmetric and converging dynamic testing algorithm is implemented, we have
\begin{align}
     P(U_i (t) = 0) \approx (1-p)(1-q)^{n\sum \limits_{j = 0}^{t-1}P(U_1 (j) = 1)}
 \end{align}
\end{lemma}

\begin{Proof}
Conditioned on $\lambda (t-1)$, we have the following recursive relation for $P(U_i(t)=0)$
\begin{align}
    P(U_i(t)=0) &= E[P(U_i(t)=0|\lambda (t-1))] \\
                &= E[P(U_i(t-1)=0|\lambda (t-1))(1-q)^{\lambda (t-1)}]\\
                &\approx E[P(U_i(t-1)=0|\lambda (t-1))]E[(1-q)^{\lambda (t-1)}] \label{lem3l1}\\
                &= P(U_i(t-1)=0)E[(1-q)^{\lambda (t-1)}] \\
                & \approx P(U_i(t-1)=0)(1-q)^{E[\lambda (t-1)]} \label{lem3l2}\\
                &= P(U_i(t-1)=0)(1-q)^{\sum \limits_{j=1}^n P(U_j(t-1)=1)} \\
                &= P(U_i(t-1)=0)(1-q)^{n P(U_1(t-1)=1)} \label{lem3l3}
\end{align}
where \eqref{lem3l1} follows from Lemma~\ref{lemma33}, \eqref{lem3l2} follows from the linear approximation of the function $(1-q)^{x} \approx 1-qx$, and \eqref{lem3l3} follows from the symmetry criterion of the implemented algorithm. Recursively using the result in \eqref{lem3l3} and the initial value $P(U_i(0)=0)=(1-p)$ yields the desired result.
\end{Proof}

In order to complete our analysis and give a lower bound for the expected number of susceptible individuals when the infection is under control, we further need to focus on $P(U_i(t)=1)$. Similar to the case of $P(U_i(t)=0)$, a direct calculation is not analytically tractable, however, we have a recursive relation when conditioned on $\lambda (t-1)$, and we obtain the following lemma.

\begin{lemma} \label{lemma4}
When a symmetric and converging dynamic testing algorithm is implemented and $cov\left(P(U_i(t)=0|\lambda (t)),p'_{\lambda (t)}(t+1)\right)$ and $cov\left(P(U_i(t)=1|\lambda (t)),p'_{\lambda (t)}(t+1)\right)$ are arbitrarily small for all $t \geq 0$, we have
\begin{align}
    P(U_i (t) = 1) \approx p((1+nq(1-p)))^t \prod \limits_{j=1}^{t}p'(j)
\end{align}
where the conditional probability of an individual not being identified in the tests at time $t$ given $\lambda (t-1)$ is denoted by $p'_{\lambda (t-1)}$.
\end{lemma}

\begin{Proof}
Conditioned on $\lambda (t-1)$, we have the following recursive relation
\begin{align}
    P(U_i & (t) =1) \nonumber\\
    =& E[P(U_i(t)=1 | \lambda (t-1))]\label{lem5l1}\\
    =& E[P(U_i(t-1)=0 | \lambda(t-1))(1-(1-q)^{\lambda (t-1)})p'_{\lambda(t-1)}(t)\nonumber\\ 
    &+P(U_i(t-1)=1 | \lambda(t-1))p'_{\lambda(t-1)}(t)] \label{lem5l2}\\ 
    \approx & E[1-(1-q)^{\lambda (t-1)}]E[P(U_i(t-1)=0 | \lambda(t-1))p'_{\lambda(t-1)}(t)]\nonumber\\ 
    &+E[P(U_i(t-1)=1 | \lambda(t-1))p'_{\lambda(t-1)}(t)] \label{lem5l3}\\ 
    \approx & E[p'_{\lambda(t-1)}(t)]E[1-(1-q)^{\lambda (t-1)}]E[P(U_i(t-1)=0 | \lambda(t-1))]\nonumber\\ 
    &+E[p'_{\lambda(t-1)}(t)]E[P(U_i(t-1)=1 | \lambda(t-1))] \label{lem5l4}\\ 
    =& p'(t)\left(P(U_i(t-1)=0)(1-E[(1-q)^{\lambda (t-1)}])+P(U_i(t-1)=1)\right)\label{lem5l5}\\
    \approx &p'(t)\left((1-p)(1-q)^{n\sum \limits_{j = 0}^{t-2}P(U_i(j) = 1)}(1-(1-q)^{E[\lambda (t-1)]})+P(U_i(t-1)=1)\right)\label{lem5l6}\\
    =& p'(t)\left((1-p)(1-q)^{n\sum \limits_{j = 0}^{t-2}P(U_i(j) = 1)}(1-(1-q)^{nP(U_i(t-1) = 1)})+P(U_i(t-1)=1)\right)\label{lem5l7}\\
    =& p'(t)\left((1-p)((1-q)^{n\sum \limits_{j = 0}^{t-2}P(U_i(j) = 1)}-(1-q)^{n\sum \limits_{j = 0}^{t-1}P(U_i(j) = 1)})+P(U_i(t-1)=1)\right)\label{lem5l8}\\
    \approx & p'(t)\left((1-p)(qn\sum \limits_{j = 0}^{t-1}P(U_i(j) = 1)-qn\sum \limits_{j = 0}^{t-2}P(U_i(j) = 1))+P(U_i(t-1)=1)\right)\label{lem5l9}\\
    =& p'(t)\left(nq(1-p)P(U_i(t-1)=1) + P(U_i(t-1)=1) \right) \\
    =& p'(t)\left(1+nq(1-p)\right) P(U_i(t-1)=1)   \label{lem5l10}
\end{align}
where \eqref{lem5l3} follows from the arbitrarily small variance of $(1-q)^{\lambda (t)}$ similar to the proof of Lemma~\ref{lemma33}, \eqref{lem5l4} follows from the given vanishing covariance assumptions in the statement of the lemma, \eqref{lem5l6} follows from Lemma~\ref{lemma3}, and \eqref{lem5l9} follows from the linear approximation $(1-q)^x \approx 1-qx$. Recursively applying \eqref{lem5l10} yields the desired result.
\end{Proof}

Combining the results of Lemma~\ref{lemma3} and Lemma~\ref{lemma4}, we have the following result.

\begin{theorem} \label{theorem1}
When a symmetric and converging dynamic testing algorithm is implemented and vanishing covariance constraints in Lemma~\ref{lemma4} are satisfied for all $t \geq 0$, we have
\begin{align}
    E[\alpha (t)] \approx n(1-p)(1-q)^{np\sum \limits_{i = 0}^{t-1}\left((1+nq(1-p))^i \prod \limits _{j=1}^{i}p'(j)\right)} \label{thm1}
\end{align}
\end{theorem}

\begin{Proof}
Expressing $\alpha (t)$ in terms of the corresponding indicator random variables and using the symmetry criterion and results of Lemma~\ref{lemma3} and Lemma~\ref{lemma4} yields 
\begin{align}
    E[\alpha (t)] &= E\left[\sum \limits_{i = 1}^{n}\mathbbm{1}_{\{U_i(t)=0\}}\right] \\
    &= \sum \limits_{i = 1}^{n} E[\mathbbm{1}_{\{U_i(t)=0\}}] \\
    &= \sum \limits_{i = 1}^{n} P(U_i(t)=0)\\
    &= n P(U_i(t)=0)\\
    &\approx n(1-p)(1-q)^{np\sum \limits_{i = 0}^{t-1}\left((1+nq(1-p))^i \prod \limits _{j=1}^{i}p'(j)\right)}
\end{align}
which is the desired result.
\end{Proof}

Our main result Theorem~\ref{theorem1} is a general result and holds for the symmetric and converging dynamic testing algorithms as long as they satisfy the vanishing covariance conditions that we state in Lemma~\ref{lemma4}. In the remainder of this section, we propose and describe two dynamic testing algorithms and analyze their performance.

\subsection{Dynamic Individual Testing Algorithm} \label{subsind}
In \emph{dynamic individual testing algorithm}, we do not utilize the group testing approach, and uniformly randomly select $T$ individuals to individually test at each time instance $t \geq 1$, from the non-isolated individuals.

In order to analyze the performance of our dynamic individual testing algorithm, we use the general result of Theorem~\ref{theorem1}. First, we show that the dynamic individual testing algorithm satisfies the symmetry and convergence criteria in \eqref{symalgcr} and \eqref{convalgcr}.

Since the process of selection of individuals to be tested is repeated at each time instance with uniformly random selections, as well as the infection spread process, dynamic individual testing algorithm is symmetric. We show that dynamic individual testing algorithm also satisfies the convergence criterion \eqref{convalgcr} in the following lemma. For the range of the testing capacity $T$, we focus on the case of $T<n$, since when $T\geq n$, at one time instance, everyone can be tested individually and the infection will be under control trivially.

\begin{lemma} \label{lemma66}
For constant $T$ and $n$, dynamic individual testing algorithm satisfies the convergence criterion
\begin{align}
        \lim_{t \rightarrow \infty}P(U_i(t)=1) = 0, \quad i \in [n]
\end{align}
\end{lemma}

\begin{Proof}
First, the probability that an infected individual is detected at a time instance $t$, denoted by $1- p'(t)$ is
\begin{align}
    1 - p'(t) &= E[1-p'_{\gamma(t-1)}(t)]\\
    &=E \left [\frac{T}{n- \gamma (t-1)} \right] \\
    &\geq \frac{T}{n}
\end{align}
where $p'_{\gamma(t-1)}(t)$ denotes the probability of the conditional event that an infected individual is not detected at the time instance $t$ given $\gamma (t-1)$. Now, since the conditional events of detection given that the individual is infected are independent across time due to the uniform random selection of tested individuals at each time instance, and the fact that
\begin{align}
    \sum \limits_{i = 1}^\infty (1-p'(i)) \geq \sum \limits_{i = 1}^\infty \frac{T}{n} \label{lem6l1}
\end{align}
since the right hand side of \eqref{lem6l1} grows to infinity, from the second Borel-Cantelli lemma, the conditional detection event occurs infinitely often, i.e., let $D_t$ denote the event that the individual $i$ is identified at time $t$, then
\begin{align}
    P(\limsup _{t \rightarrow \infty} D_t) = 1
\end{align}
which yields the desired result of $\lim_{t \rightarrow \infty}P(U_i(t)=1) = 0$.
\end{Proof}

Next, we consider a weak version of our algorithm, where at each time instance, during the testing phase, instead of selecting $T$ individuals to test from $n-\gamma (t)$ non-isolated individuals, we select $T$ individuals from $n$ individuals, including the isolated ones, whose test results will be negative. For the weak version of the dynamic individual testing algorithm, we have
\begin{align}
    1-p'(t) = \frac{T}{n}, \quad t > 0
\end{align}
which is the identification probability of an individual at time $t$. Moreover, since it is an upper bound for the identification probability of an individual for the original dynamic individual testing algorithm, we have
\begin{align}
    \lim_{t \rightarrow \infty}E[\alpha_{orig}(t)] \geq \lim_{t \rightarrow \infty}E[\alpha_{weak}(t)]
\end{align}
Since the weak dynamic individual testing algorithm is a symmetric and converging algorithm (note that the result of Lemma~\ref{lemma66} still holds) and due to the fact that $p'(t)$ is constant in the weak dynamic individual testing algorithm, we can directly use the result of Lemma~\ref{lemma4}, due to the fact that the vanishing covariance criteria are already satisfied. Now, using Theorem~\ref{theorem1}, we have the following result for the weak dynamic individual testing algorithm.

\begin{theorem} \label{theorem2}
When weak dynamic individual testing algorithm is used and $(1-\frac{T}{n})(1+nq(1-p))<1$, we have
\begin{align}
    \lim_{t \rightarrow \infty} E[\alpha_{weak} (t)] \approx n(1-p)(1-q)^{\frac{np}{1-(1-\frac{T}{n})(1+nq(1-p))}}
\end{align}
which is a lower bound for $\lim_{t \rightarrow \infty} E[\alpha _ {orig} (t)]$, i.e., the limit of the expected number of susceptible individuals for the dynamic individual testing algorithm.
\end{theorem}

\begin{Proof}
The weak dynamic individual testing algorithm satisfies the constraints for using Theorem~\ref{theorem1}. Thus, we can use Theorem~\ref{theorem1} directly to derive the long term behavior of the expected number of the susceptible individuals by considering the limit of \eqref{thm1} for constant $p'(t) = 1-T/n$. On the other hand, in the case of $(1-\frac{T}{n})(1+nq(1-p)) \geq 1$, we have $\lim_{t \rightarrow \infty} E[\alpha_{weak} (t)] \approx 0$.
\end{Proof}

\subsection{Dynamic Dorfman Type Group Testing Algorithm} \label{subsdorf}
In \emph{dynamic Dorfman type group testing algorithm}, we utilize the group testing idea while designing the test matrices at each time instance $t \geq 1$.

At each time instance, dynamic Dorfman type group testing algorithm uniformly randomly partitions the set of all non-isolated individuals to equal sized $T/2$ disjoint sets (with possible 1 unequal sized set if the total number of non-isolated individuals is not divisible by $T/2$). Then, test samples of the individuals are mixed with others in the same group: $T/2$ group tests are performed, and positive and negative groups are determined. Then, among the positive groups, one group (or multiple groups if the sizes of the groups are less than $T/2$, depending on the system parameters) is uniformly randomly selected to be individually tested. $T/2$ individuals from the selected group are uniformly randomly selected and individually tested; here depending on the parameters, some individuals from the selected group may not be tested, as well as individuals from multiple positive groups may be selected. Detected infections are isolated and at the next time instance, the whole process is repeated with uniform random selections.

Since the partition selection and individuals within group selection are uniformly random at each time instance, dynamic Dorfman type group testing algorithm is symmetric. Similar to Section~\ref{subsind}, we proceed by showing that the dynamic Dorfman type group testing algorithm satisfies the convergence criterion in \eqref{convalgcr} as well.

\begin{lemma} \label{lemma7}
For constant $T$ and $n$, dynamic Dorfman testing algorithm satisfies the convergence criterion
\begin{align}
        \lim_{t \rightarrow \infty}P(U_i(t)=1) = 0, \quad i \in [n]
\end{align}
\end{lemma}

\begin{Proof}
The probability that an individual is identified at a time instance $t$, which is $1- p'(t)$, satisfies the following
\begin{align}
    1 - p'(t) &\geq \frac{T}{2n}
\end{align}
since $T/2$ individuals are individually tested at each time instance and due to the symmetry of the infection status in the system and the fact that the individuals are selected from a positive group (or from multiple positive groups), the probability of detection for the dynamic Dorfman type group testing algorithm, at each time instance, must be higher than uniformly randomized testing of $T/2$ individuals. Now, since the events of identification of individuals are independent across time due to the uniform random selection of tested individuals at each time instance, and the fact that
\begin{align}
    \sum \limits_{i = 1}^\infty (1-p'(t)) &\geq \sum \limits_{i = 1}^\infty \frac{T}{2n} \label{lem7l1}
\end{align}
grows to infinity, we conclude that $\lim_{t \rightarrow \infty}P(U_i(t)=1) = 0$, from the second Borel-Cantelli lemma as in Lemma~\ref{lemma66}.
\end{Proof}
Similar to the dynamic individual testing case, we focus on a weak version of the dynamic Dorfman type group testing algorithm to provide a lower bound for the expected number of susceptible individuals in the system at the steady state. 

In the weak version of the dynamic Dorfman type group testing algorithm, the results from the $T/2$ group tests are discarded and it is basically equivalent to the uniformly random individual testing of $T/2$ individuals. Furthermore, the isolated individuals are also included in the testing procedure: $n$ individuals are divided into groups and then tested at each time instance, rather than only non-isolated individuals as in the original dynamic Dorfman type group testing algorithm. The probability of identification at time $t$ for the weak dynamic Dorfman type group testing algorithm, given by $1-p'(t)$, is always less than the original dynamic Dorfman type group testing algorithm, due to the discarded $T/2$ group tests and included isolated individuals to the tests. Note that the weak dynamic Dorfman type group testing algorithm is also symmetric and satisfies the convergent criterion \eqref{convalgcr}, since Lemma~\ref{lemma7} still holds; the lower bound in \eqref{lem7l1} is the detection probability of the weak algorithm. Moreover, since the weak algorithm has constant value for $p'_{\lambda (t-1)}(t)$, it satisfies the vanishing covariance constraints given in the statement of Lemma~\ref{lemma4}. Using the general result of Theorem~\ref{theorem1}, we have the following result for dynamic Dorfman type group testing algorithm by following similar steps to those in Theorem~\ref{theorem2}.

\begin{theorem} \label{theorem3}
When weak dynamic Dorfman type group testing algorithm is used and $(1-\frac{T}{2n})(1+nq(1-p))<1$, we have
\begin{align}
    \lim_{t \rightarrow \infty} E[\alpha_{weak} (t)] \approx n(1-p)(1-q)^{\frac{np}{1-(1-\frac{T}{2n})(1+nq(1-p))}}
\end{align}
which is a lower bound for $\lim_{t \rightarrow \infty} E[\alpha _ {orig} (t)]$, i.e., the expected number of susceptible individuals for the dynamic Dorfman type group testing algorithm.
\end{theorem}

Note that, this result of weak dynamic Dorfman type group testing algorithm is a loose lower bound for the performance of the algorithm, which is only significant because it shows that, the weak dynamic Dorfman type group testing algorithm performs in a similar manner with weak dynamic individual testing algorithm, order-wise ($T$ replaced with $T/2$), which is a performance lower bound for the dynamic Dorfman type group testing algorithm.

\subsection{Comparison of Dynamic Individual and Dorfman Algorithms} 

To compare the average number of detected infections at a given time instance for the dynamic individual testing and dynamic Dorfman type group testing algorithms, we obtain the following results stated in the following lemmas.

\begin{lemma} \label{lem8}
When there are $\tilde{\alpha} (t)$ susceptible and $\tilde{\lambda} (t)$ non-isolated infected individuals in a system after the infection spread phase and just before the testing phase at time instance $t$, and dynamic individual testing algorithm is being used, on average, $\frac{T\tilde{\lambda}(t)}{\tilde{\alpha}(t)+\tilde{\lambda}(t)}$ infections are detected and isolated at time $t$.
\end{lemma}

\begin{Proof}
When $T$ individuals from $\tilde{\alpha}(t)+\tilde{\lambda}(t)$ individuals are uniformly randomly selected, we have
\begin{align}
    E\left[\sum\limits_{i=1}^{T} \mathbbm{1}_{\tilde{U}_i (t) = 1}\right] &= TP(\tilde{U}_i (t) = 1) \\
    &=\frac{T\tilde{\lambda}(t)}{\tilde{\alpha}(t)+\tilde{\lambda}(t)}
\end{align}
where $\tilde{U}_i(t)$ represents the infection status of the $i$th selected individual for testing at the time of testing phase.
\end{Proof}

On the other hand, when dynamic Dorfman type group testing algorithm is used, $T/2$ individuals to be individually tested are chosen from a set of individuals of size $\frac{2(\tilde{\alpha}(t)+\tilde{\lambda}(t))}{T}$ that is guaranteed to have at least one infected individual, in the case of $\tilde{\alpha}(t)+\tilde{\lambda}(t) \geq T^2/4$. When $\tilde{\alpha}(t)+\tilde{\lambda}(t) < T^2/4$, $T/2$ individuals to be tested individually are chosen from multiple groups, each having at least one infected individual. The following lemma gives an average number of detected and isolated infections at each time instance for the case of $\tilde{\alpha}(t)+\tilde{\lambda}(t) \geq T^2/4$, which, in general, holds for practical applications with low testing capacity. Moreover, the following result is also a lower bound for the case of $\tilde{\alpha}(t)+\tilde{\lambda}(t) < T^2/4$, where $T/2$ individuals to be individually tested are selected from multiple positive groups.

\begin{lemma} \label{lem9}
When there are $\tilde{\alpha} (t)$ susceptible and $\tilde{\lambda} (t)$ non-isolated infected individuals in a system after the infection spread phase and just before the testing phase at time instance $t$, with $\tilde{\alpha}(t)+\tilde{\lambda}(t) \geq T^2/4$, and dynamic Dorfman type group testing algorithm is being used, if $\tilde{\alpha}(t) \geq 2(\tilde{\alpha}(t)+\tilde{\lambda}(t))/T$, on average,
\begin{align}
    \frac{T\tilde{\lambda}(t)}{2(\tilde{\alpha}(t)+\tilde{\lambda}(t))} \left(1-\frac{\binom{\tilde{\alpha}(t)}{2(\tilde{\alpha}(t)+\tilde{\lambda}(t))/{T}}}{\binom{\tilde{\alpha}(t) + \tilde{\lambda}(t)}{2(\tilde{\alpha}(t)+\tilde{\lambda}(t))/{T}}}\right)^{-1} \label{lem9res}
\end{align}
infections are detected and isolated at time $t$. If $\tilde{\alpha}(t) < 2(\tilde{\alpha}(t)+\tilde{\lambda}(t))/T$, then, on average, $\frac{T\tilde{\lambda}(t)}{2(\tilde{\alpha}(t)+\tilde{\lambda}(t))}$ infections are detected and isolated at time $t$. In the case of $\tilde{\alpha}(t)+\tilde{\lambda}(t) < T^2/4$, \eqref{lem9res} is a lower bound for the average number of detected and isolated individuals at time $t$.
\end{lemma}

\begin{Proof}
When $T/2$ individuals are uniformly randomly selected from a set of individuals that are guaranteed to have at least one infection, with size $2(\tilde{\alpha}(t)+\tilde{\lambda}(t))/{T}$, we have
\begin{align}
    E\left[\sum\limits_{i=1}^{T/2} \mathbbm{1}_{\tilde{U}_i(t) = 1} \middle|\; \sum\limits_{i=1}^{C} \mathbbm{1}_{\tilde{U}_i(t) = 1} \geq 1\right] &= \frac{E\left[\sum\limits_{i=1}^{T/2} \mathbbm{1}_{\tilde{U}_i(t) = 1}\right]}{P\left(\sum\limits_{i=1}^{C} \mathbbm{1}_{\tilde{U}_i(t) = 1} \geq 1\right)} \label{lem9l1}\\
    &=    \frac{T\tilde{\lambda}(t)}{2(\tilde{\alpha}(t)+\tilde{\lambda}(t))} \left(1-\frac{\binom{\tilde{\alpha}(t)}{2(\tilde{\alpha}(t)+\tilde{\lambda}(t))/{T}}}{\binom{\tilde{\alpha}(t) + \tilde{\lambda}(t)}{2(\tilde{\alpha}(t)+\tilde{\lambda}(t))/{T}}}\right)^{-1}
\end{align}
where $\tilde{U}_i(t)$ represents the infection status of the $i$th selected individual for testing, at the time of testing phase and $C = 2(\tilde{\alpha}(t)+\tilde{\lambda}(t))/{T}$.

In the case of $\tilde{\alpha}(t)+\tilde{\lambda}(t) < T^2/4$, $T/2$ individuals to be tested individually are chosen from multiple groups where each of them is guaranteed to have at least one infected individual. Therefore, the term in the denominator of the right hand side of \eqref{lem9l1}, i.e. $P\left(\sum\limits_{i=1}^{C} \mathbbm{1}_{\tilde{U}_i(t) = 1} \geq 1\right)$, is replaced by the probability of the event that multiple subsets of size $C$ having at least one non-isolated infected member, which is a subset of the event that only one subset of individuals of size $C$ having at least one non-isolated infected member and thus, having lower probability. Therefore, \eqref{lem9l1} is also a lower bound for the average number of detected and isolated infections at time instance $t$, for the case of $\tilde{\alpha}(t)+\tilde{\lambda}(t) < T^2/4$.
\end{Proof}

For a given state of the system at the time of the testing phase, i.e., $\tilde{\alpha}(t)$ and $\tilde{\lambda}(t)$, as we show in Lemma~\ref{lem8} and Lemma~\ref{lem9}, using dynamic Dorfman type group testing algorithm becomes advantageous with respect to the dynamic individual testing algorithm when $\tilde{\alpha}(t) \geq 2(\tilde{\alpha}(t)+\tilde{\lambda}(t))/T$ and
\begin{align}
    1/2 &< \frac{\binom{\tilde{\alpha}(t)}{2(\tilde{\alpha}(t)+\tilde{\lambda}(t))/{T}}}{\binom{\tilde{\alpha}(t) + \tilde{\lambda}(t)}{2(\tilde{\alpha}(t)+\tilde{\lambda}(t))/{T}}} \\
    &= \frac{\prod\limits_{i=0}^{C}(\tilde{\alpha}(t)-i)}{\prod\limits_{i=0}^{C}(\tilde{\alpha}(t)+\tilde{\lambda}(t)-i)}
\end{align}
where $C = 2(\tilde{\alpha}(t)+\tilde{\lambda}(t))/{T}$. 

In the next section, we present numerical results of our two proposed dynamic algorithms, as well as their weak versions, under various sets of system parameters.

\section{Numerical Results}
\begin{figure} [t]
	\centering
	\begin{subfigure}[b]{0.475\textwidth}
		\centering
		\epsfig{file=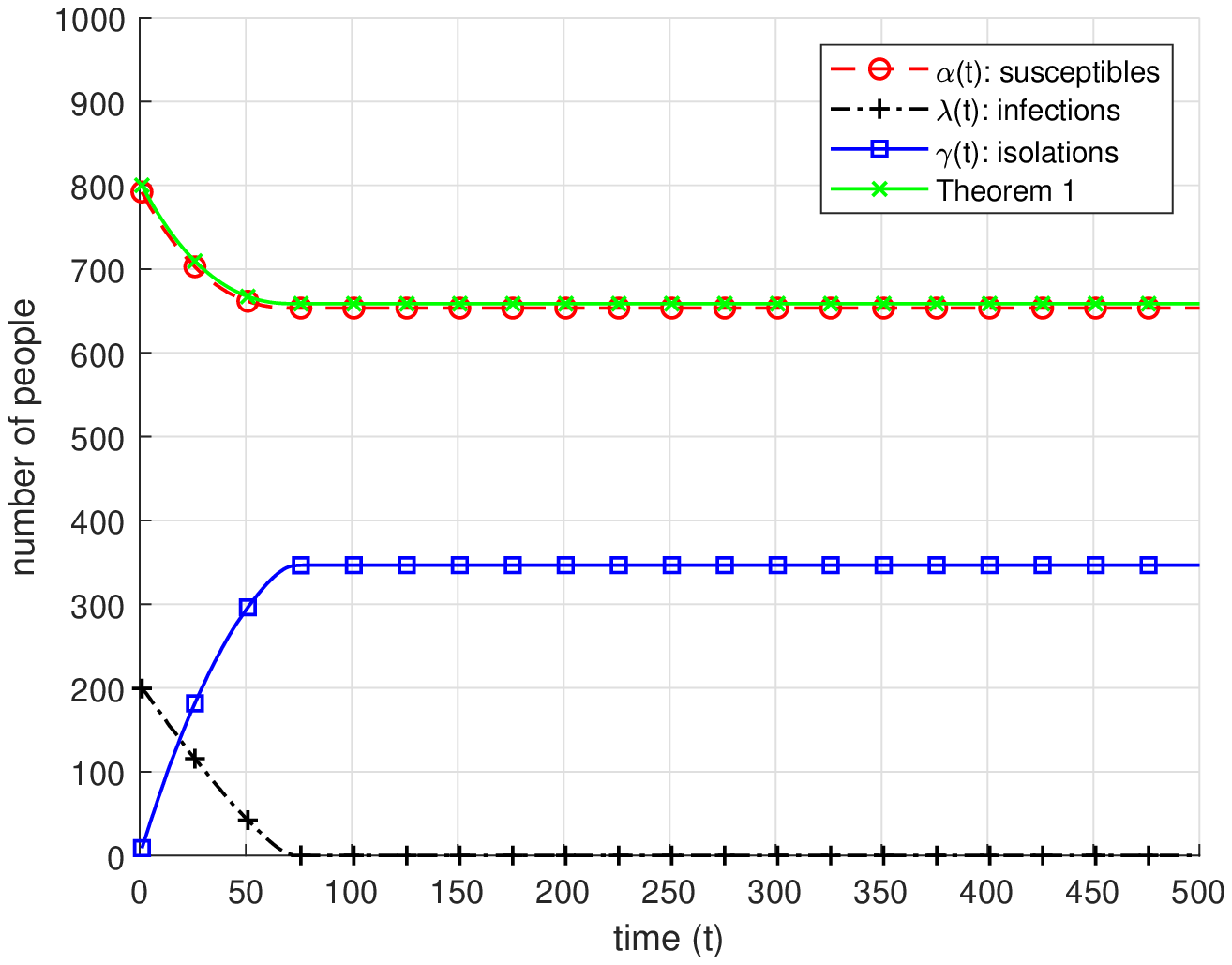,scale = 0.54} 
		\caption[]%
        {}
	\end{subfigure}
	\hfill
	\begin{subfigure}[b]{0.475\textwidth}  
		\centering 
		\epsfig{file=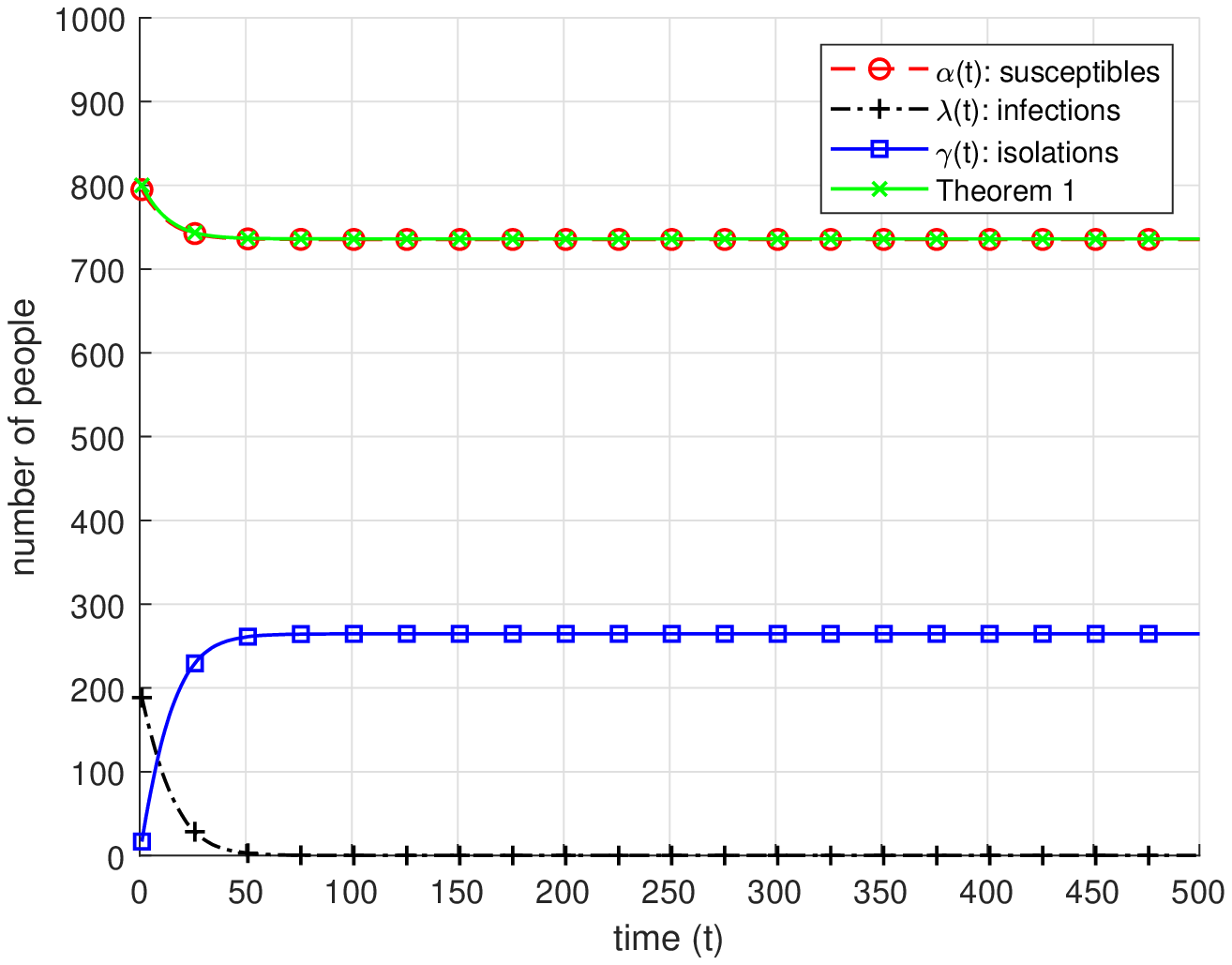,scale = 0.54}
		\caption[]%
        {}
	\end{subfigure}
	\vskip\baselineskip
	\begin{subfigure}[b]{0.475\textwidth}   
		\centering 
		\epsfig{file=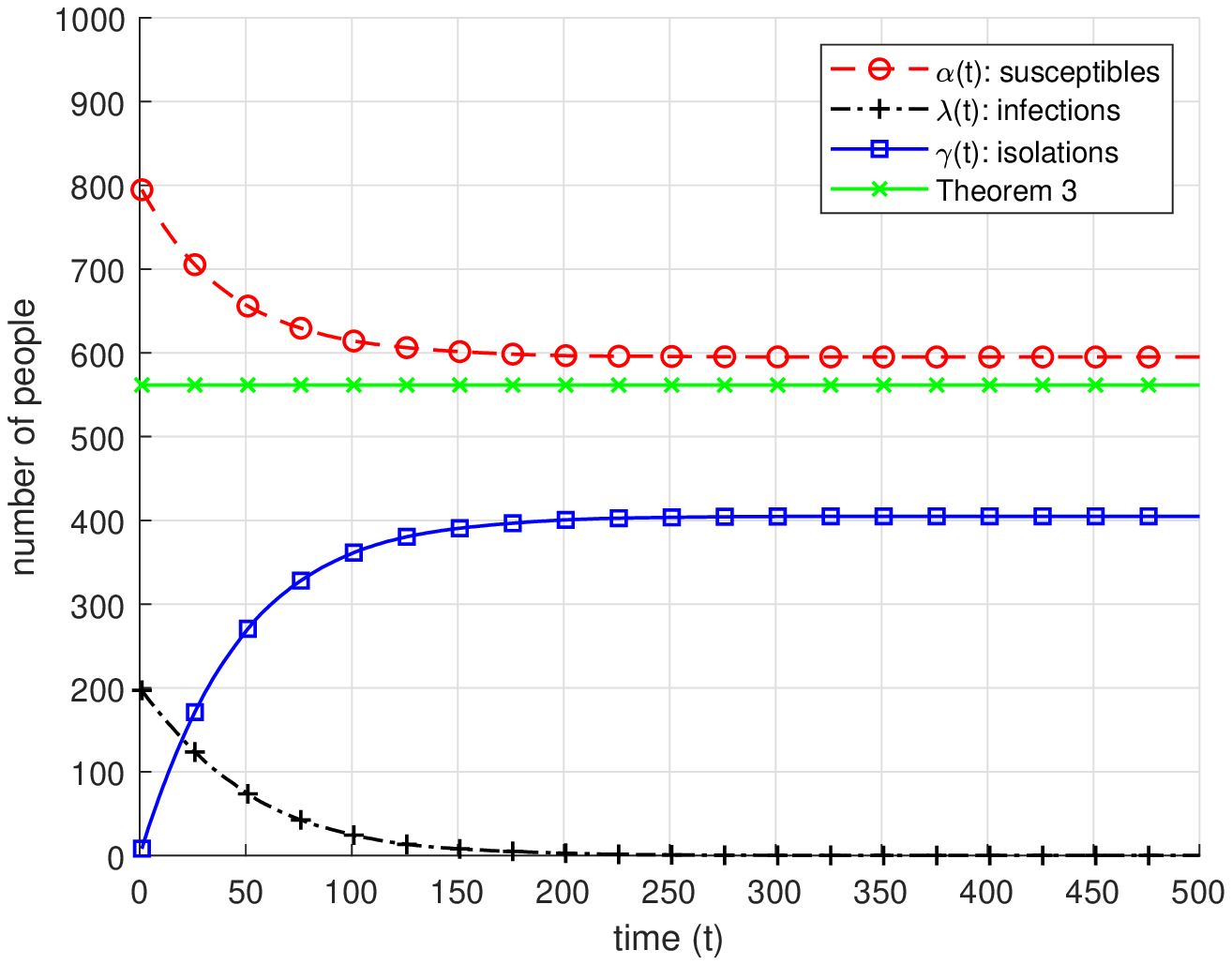,scale = 0.54}
		\caption[]%
        {}
	\end{subfigure}
	\hfill
	\begin{subfigure}[b]{0.475\textwidth}   
		\centering 
		\epsfig{file=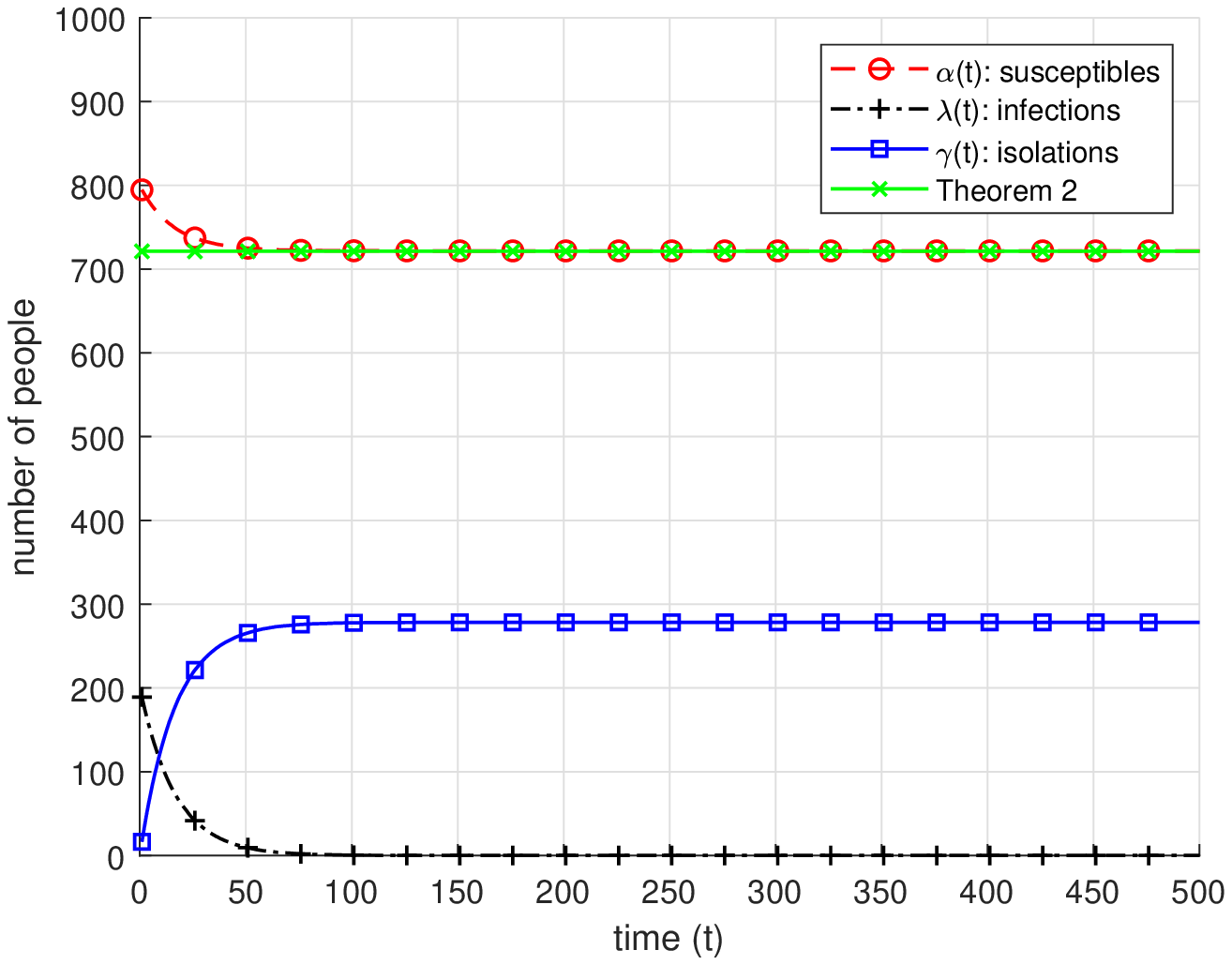,scale = 0.54}
		\caption[]%
        {}
	\end{subfigure}
	\caption{Average values of the random processes $\alpha(t)$, $\lambda(t)$ and $\gamma(t)$, with obtained theoretical approximations given in Theorem~\ref{theorem1}, Theorem~\ref{theorem2} and Theorem~\ref{theorem3} when $n=1000$, $T=80$, $q=0.00003$, $p=0.2$, for (a) dynamic Dorfman type group testing algorithm, (b) dynamic individual testing algorithm, (c) weak dynamic Dorfman type group testing algorithm, (d) weak dynamic individual testing algorithm.} 
	\label{sims1}
\end{figure}

\begin{figure} [t]
	\centering
	\begin{subfigure}[b]{0.475\textwidth}
		\centering
		\epsfig{file=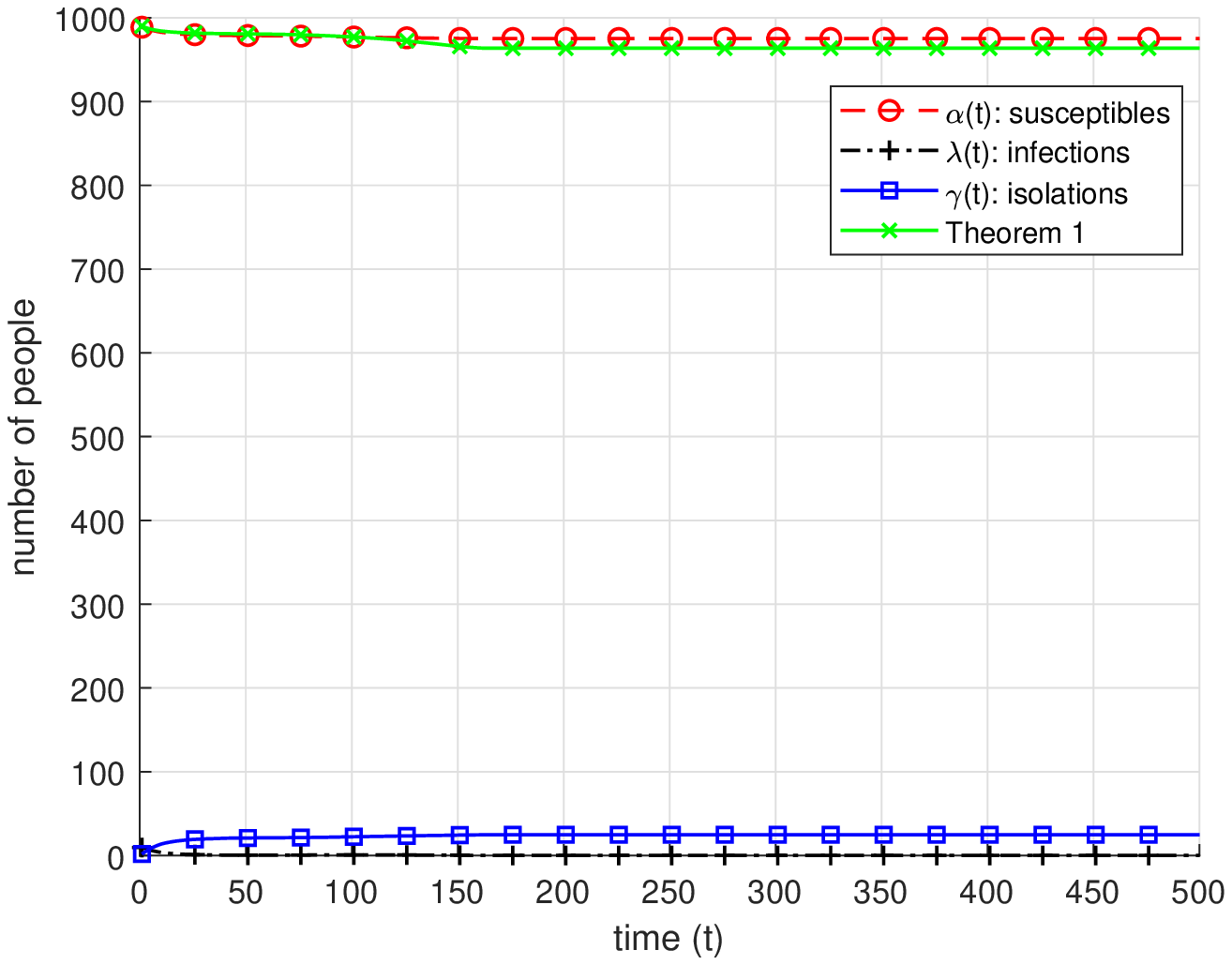,scale = 0.54} 
		\caption[]%
        {}
	\end{subfigure}
	\hfill
	\begin{subfigure}[b]{0.475\textwidth}  
		\centering 
		\epsfig{file=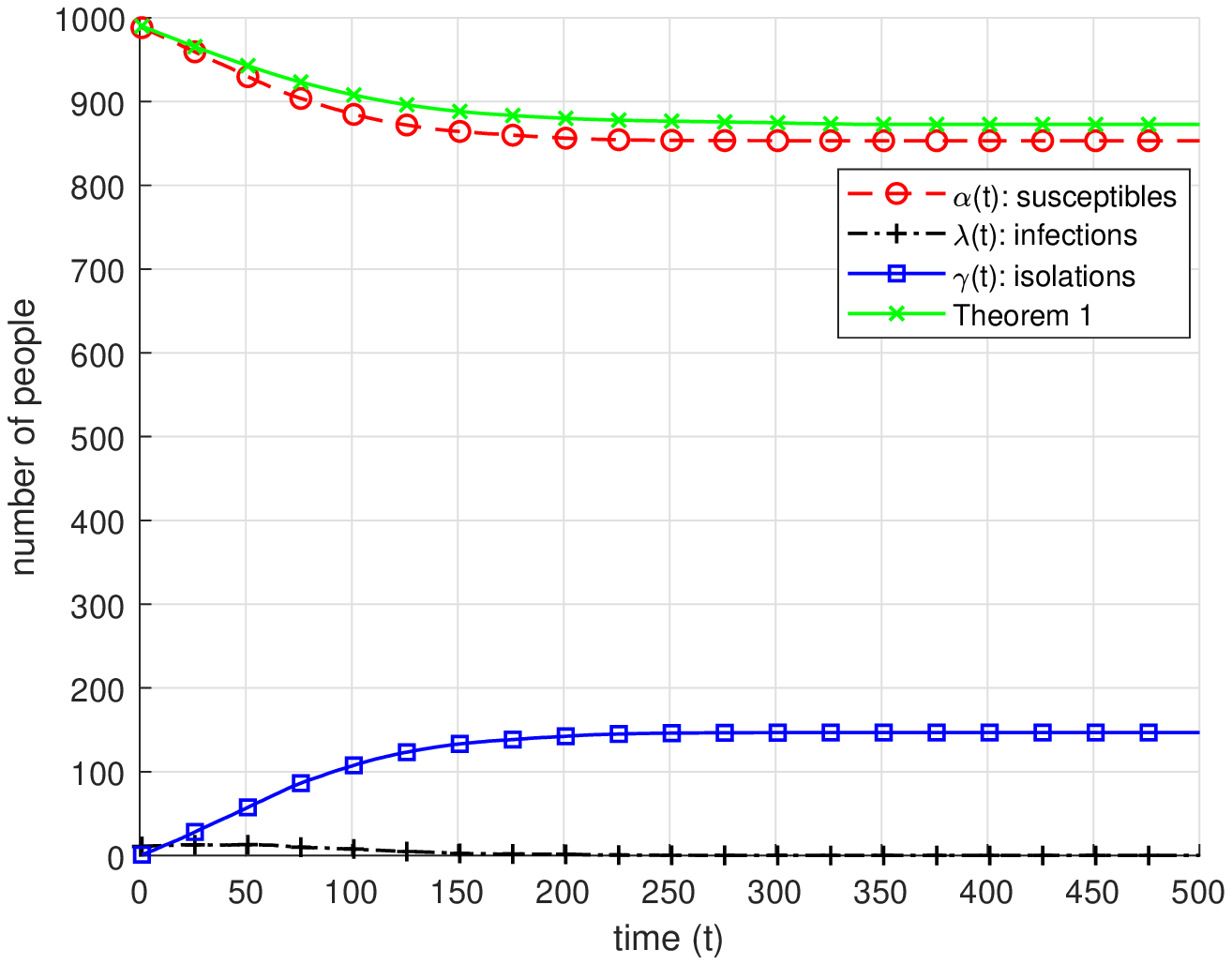,scale = 0.54}
		\caption[]%
        {}
	\end{subfigure}
	\vskip\baselineskip
	\begin{subfigure}[b]{0.475\textwidth}   
		\centering 
		\epsfig{file=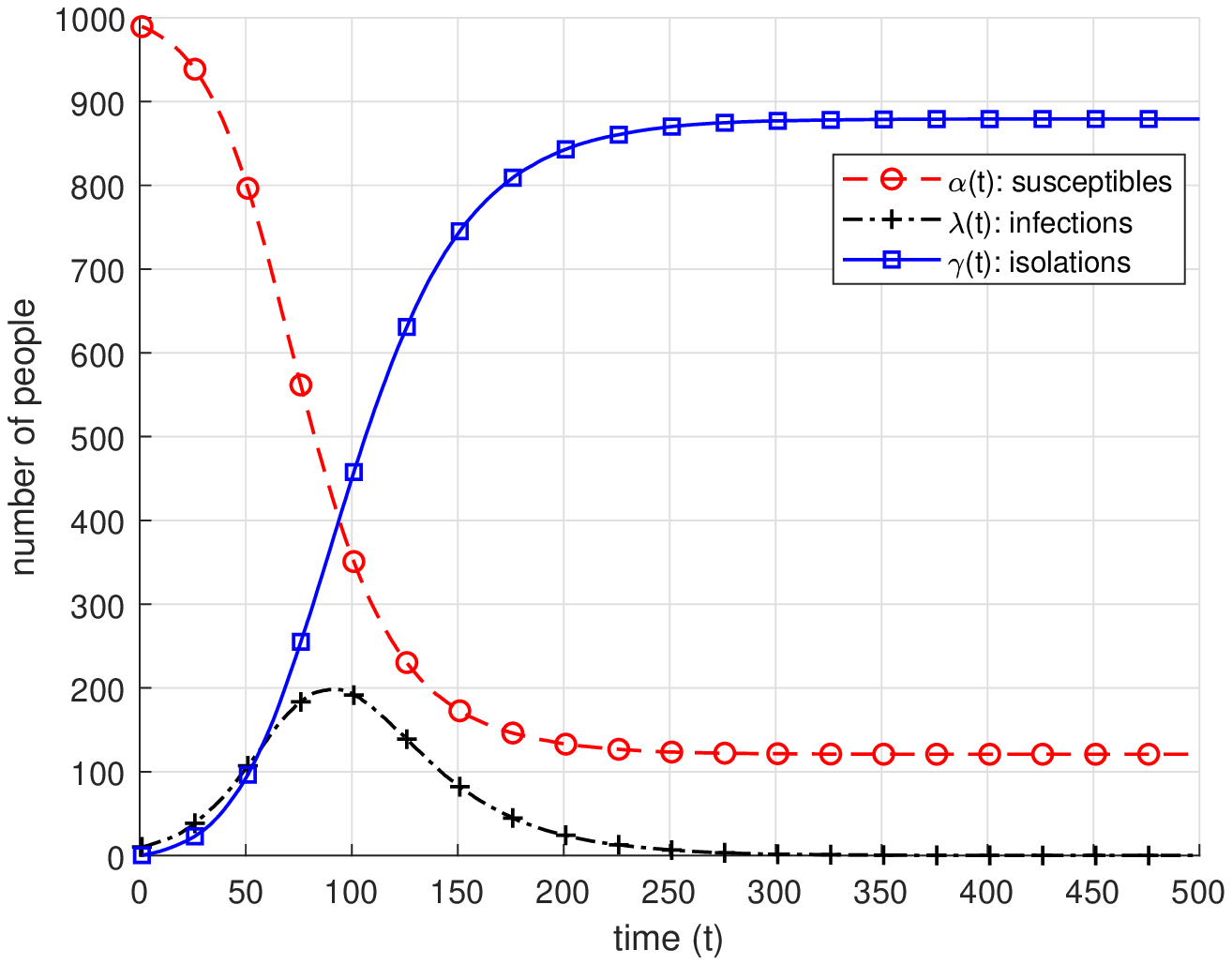,scale = 0.54}
		\caption[]%
        {}
	\end{subfigure}
	\hfill
	\begin{subfigure}[b]{0.475\textwidth}   
		\centering 
		\epsfig{file=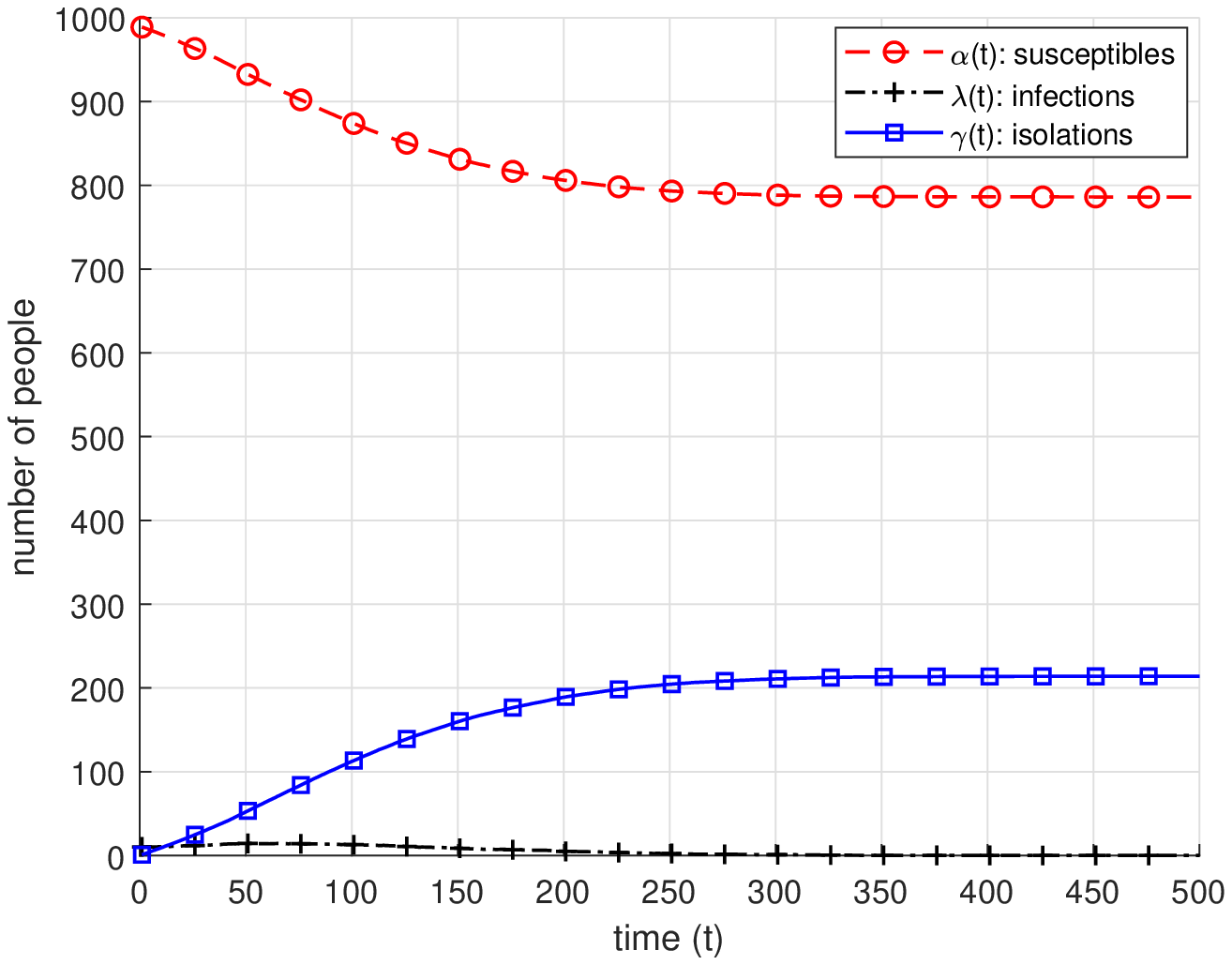,scale = 0.54}
		\caption[]%
        {}
	\end{subfigure}
    \caption{Average values of the random processes $\alpha(t)$, $\lambda(t)$ and $\gamma(t)$, with obtained theoretical approximations given in Theorem~\ref{theorem1} when $n=1000$, $T=80$, $q=0.0001$, $p=0.01$, for (a) dynamic Dorfman type group testing algorithm, (b) dynamic individual testing algorithm, (c) weak dynamic Dorfman type group testing algorithm, (d) weak dynamic individual testing algorithm.} 
	\label{sims2}
\end{figure}

\begin{figure}[t]
    \centering
    \begin{subfigure}[t]{0.475\textwidth}
        \centering
        \epsfig{file=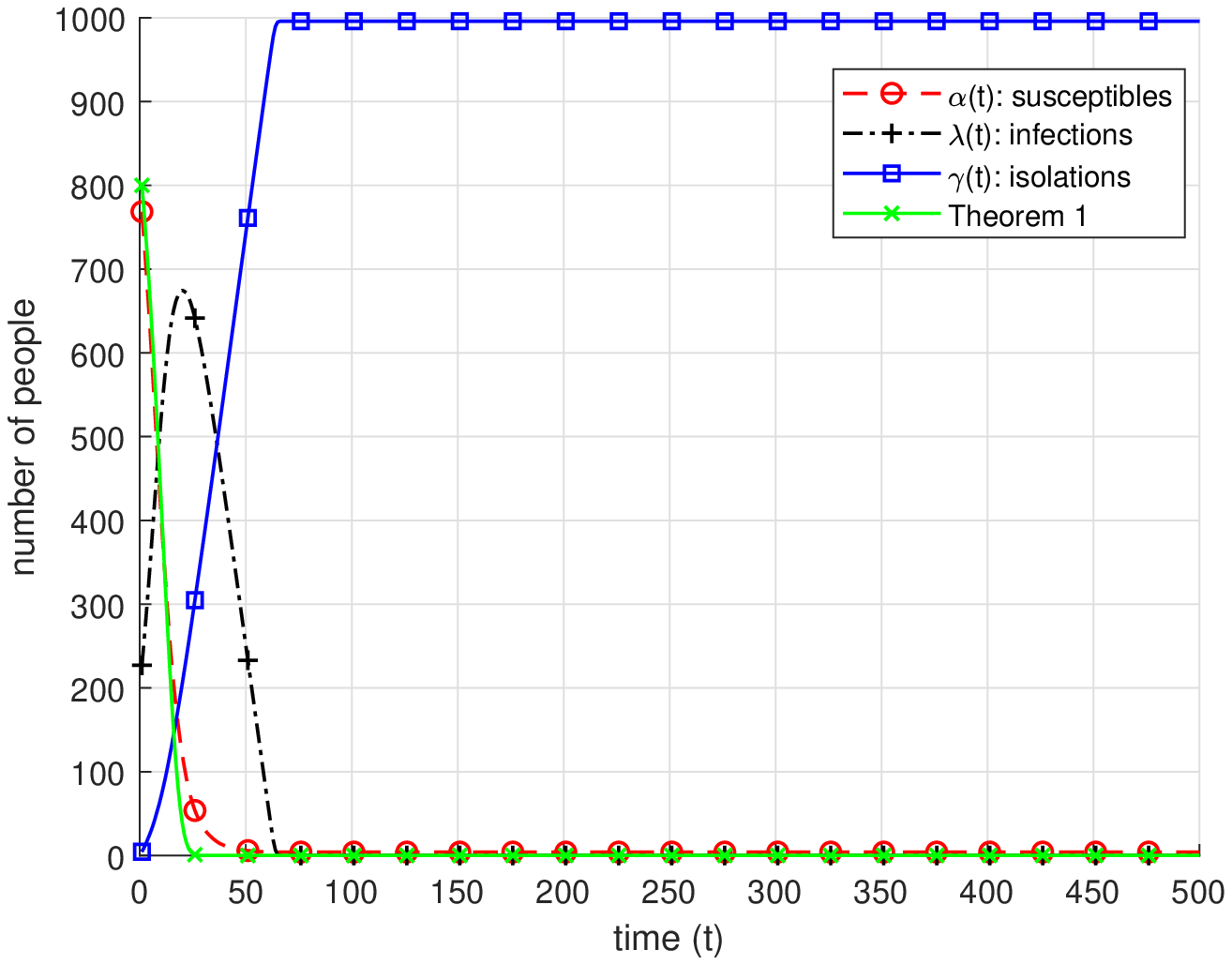,scale=0.54}
        \caption[]%
        {}
    \end{subfigure}%
    ~ 
    \begin{subfigure}[t]{0.475\textwidth}
        \centering
        \epsfig{file=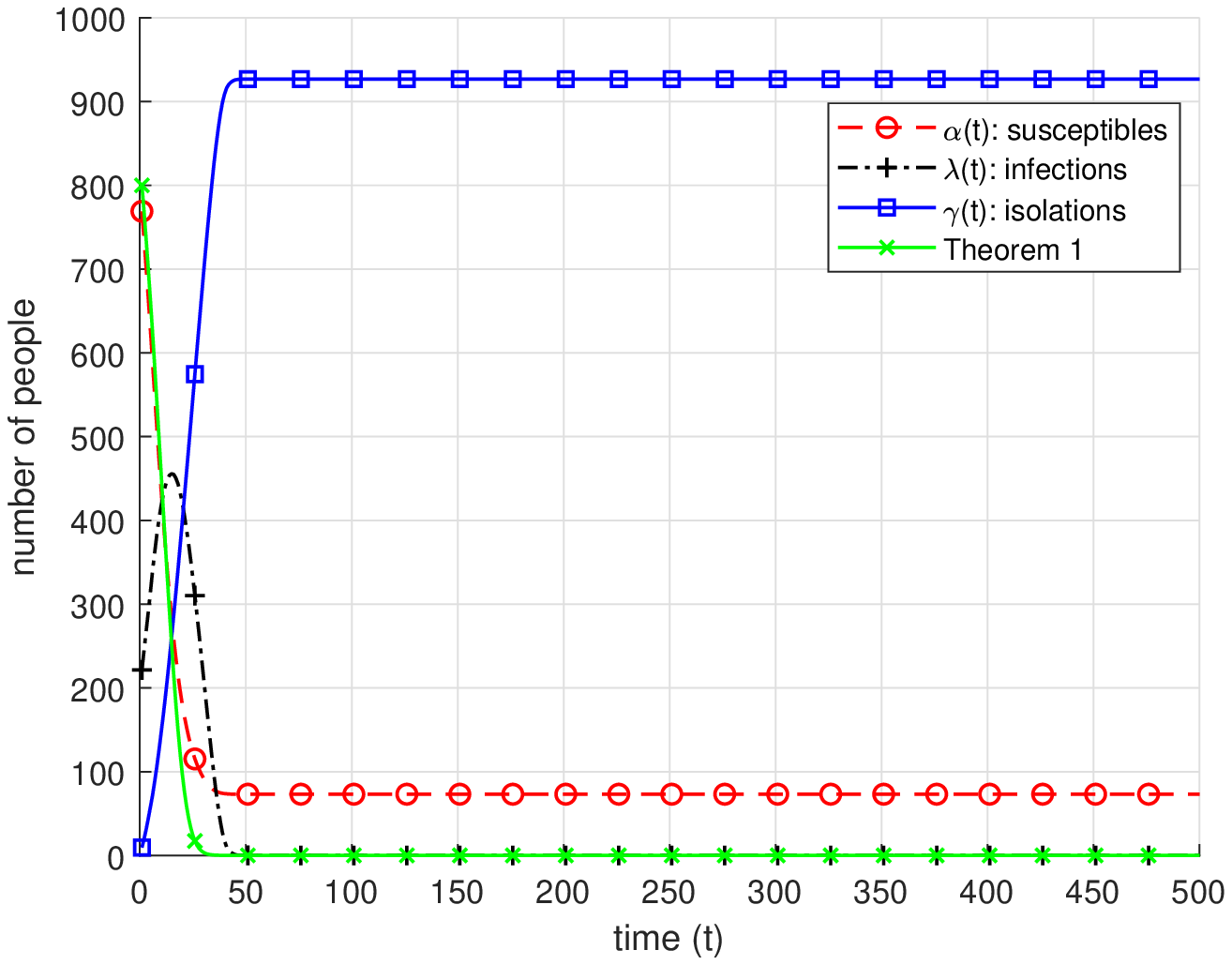,scale=0.54}
        \caption[]%
        {}
    \end{subfigure}
    \caption{Average values of the random processes $\alpha(t)$, $\lambda(t)$ and $\gamma(t)$, with obtained theoretical approximations given in Theorem~\ref{theorem1} when $n=1000$, $T=40$, $q=0.0002$, $p=0.2$, for (a) dynamic Dorfman type group testing algorithm, (b) dynamic individual testing algorithm.} 
    \label{sims3}
\end{figure}

In our numerical results, we implement the algorithms that we proposed: dynamic Dorfman type group testing algorithm, dynamic individual testing algorithm and the weak versions of these algorithms. In all of our simulations, we start with $n$ individuals with all of them susceptible. Then, at time $t=0$, we realize the initial infections in the system uniformly randomly with probability $p$. At each time instance that follows, for the infection spread phase, we simulate the random infection spread from the non-isolated infections to the susceptible individuals. For the testing phase, we simulate the random selection of individuals to be tested and perform the tests. Depending on the test results, we simulate the isolation of the detected infections. We repeat these phases at each time instance until time $t=500$, and obtain the sample paths of the random processes $\alpha(t)$, $\lambda(t)$ and $\gamma(t)$. We iterate this whole process $1000$ times to obtain $1000$ sample paths of the random processes, and then we calculate the average of the sample paths to obtain the expected values of $\alpha(t)$, $\lambda(t)$ and $\gamma(t)$, numerically. In Figure~\ref{sims1}, Figure~\ref{sims2} and Figure~\ref{sims3}, we plot these expected values of $\alpha(t)$, $\lambda(t)$ and $\gamma(t)$ for the algorithms that we propose. In our simulations, we also consider the value of the theoretical approximation result that we obtained in Theorem~\ref{theorem1}. For each sample path, at each time instance, we numerically calculate the values of $p'(t)$ for both dynamic individual testing and dynamic Dorfman type group testing algorithms, and then use the expression that we obtained in Theorem~\ref{theorem1} to calculate the $\alpha(t)$ approximation curve. We calculate and plot the average of the $\alpha(t)$ approximation curve. For the weak versions of the proposed algorithms, we use the results of Theorem~\ref{theorem2} and Theorem~\ref{theorem3} to directly calculate and plot the steady state approximations of $\alpha(t)$.

In Figure~\ref{sims1}, we present numerical results for the system with the parameters $n=1000$, $T=80$, $q=0.00003$ and $p=0.2$. Due to the relatively high number of initial infections in the system, we observe that the dynamic individual testing algorithm performs better than the dynamic Dorfman type group testing algorithm in terms of the average steady state $\alpha(t)$. In the weak versions of the algorithms, we observe that their performance is strictly worse than their respective original algorithms, at each time instance, in terms of the average $\alpha(t)$, as expected. The difference of the average $\alpha(t)$ curves between the original and weak versions of the dynamic Dorfman type group testing algorithm is higher than the difference of the average $\alpha(t)$ curves between the original and weak versions of the dynamic individual testing algorithm. This is due to the fact that, in the weak dynamic individual testing algorithm, we still utilize $T$ tests at each time instance but can sample the isolated individuals to test while in the weak dynamic Dorfman type group testing algorithm, we ignore the group tests and only consider $T/2$ individual tests. However, since the advantage of the group test is not effective for this set of parameters, as we present in Figure~\ref{sims1}, even the weak dynamic Dorfman type group testing algorithm provides a reasonable lower bound for its original version. Finally, we observe that our approximation results in Theorem~\ref{theorem1} match with the average $\alpha(t)$ curves in both dynamic Dorfman type group testing and dynamic individual testing algorithms. Similarly, the average $\alpha(t)$ curves that we obtain from the weak versions of the proposed algorithms are also closely approximated by the results that we obtain in Theorem~\ref{theorem2} and Theorem~\ref{theorem3}.

In Figure~\ref{sims2}, we run the same simulations as in Figure~\ref{sims1}, for the parameters $n=1000$, $T=80$, $q=0.0001$ and $p=0.01$. Now, relative to the first set of parameters, the number of initial infections is lower but the infection spread probability is higher. Because of the targeted individual testing to the positive groups in the dynamic Dorfman type group testing algorithm, it outperforms the dynamic individual testing algorithm for this set of parameters, as we present in Figure~\ref{sims2}. Since the advantage of the group testing is more prevalent for this set of parameters, the weak version of the dynamic Dorfman type group testing algorithm results in an average $\alpha(t)$ curve that is a loose lower bound for the average $\alpha(t)$ curve of the original version, while the weak dynamic individual testing algorithm results in a proper lower bound for the original version. Furthermore, for this set of parameters, despite the fact that the Theorem~\ref{theorem1} approximation matches the average $\alpha(t)$ curves for both of the original versions of the proposed algorithms, the resulting Theorem~\ref{theorem2} and Theorem~\ref{theorem3} approximations cannot be used due to the non-convergent exponents in the expressions.

In our third and final set of parameters, we consider lower number of test capacity, $T$, than the first two sets of parameters, high number of initial infections, $p$, and high infection spread probability, $q$. As expected, for this set of parameters, for both of the algorithms, system reaches the steady state when almost everyone in the population gets infected. Due to the high number of infections at each time instance in the system, dynamic individual testing algorithm performs slightly better than the dynamic Dorfman type group testing algorithm, even though it still fails to control the infection spread in an effective manner.

\section{Conclusions}
In this paper, we considered a dynamic infection spread model over discrete time, inspired by the SIR model, widely used in the modelling of contagious infections in populations. Instead of recovered individuals in the system, we considered isolated infections, where infected individuals can be identified and isolated via testing. In our system model, infection status of the individuals are random processes, rather than random variables such as the infection status of the individuals in the classical group testing problems. In parallel with the dynamic configuration of our system, we considered dynamic group testing algorithms: At each time instance, after the infection is spread by infected individuals to the susceptible individuals randomly, a given limited number of (possibly group) tests are performed to identify and isolate infected individuals. This dynamic infection spread and identification system is more challenging than the classical group testing problem setup, since negative identifications are not finalized and can change over time while only the positive identifications are isolated for the rest of the process. We analyzed the performance of dynamic testing algorithms by providing approximation results for the expected number of susceptible individuals (that have never gotten infected) when the infection is brought under control where all infections are identified and isolated, for symmetric and converging algorithms. Then, we proposed two dynamic algorithms: \emph{dynamic individual testing algorithm} and \emph{dynamic Dorfman type group testing algorithm}. We considered the weak versions of these algorithms and used our general result to provide lower bounds on the expected number of susceptible individuals for these two algorithms. We compared the average identification performance of these two algorithms by deriving conditions when one algorithm outperforms the other. In our simulations, we implemented both the original and weak versions of the proposed algorithms and also simulated and compared the theoretical approximation results that we derived, for three different sets of parameters and we demonstrated various possible scenarios. Our work is unique in that the disease spread in our dynamic system is due to limited testing capacity as opposed to delay in obtaining (unlimited) test results in the existing literature.

\bibliographystyle{unsrt}
\bibliography{references_grouptesting}

\end{document}